\documentclass[twocolumn,preprintnumbers,aps,prb,superscriptaddress]{revtex4-1}
\usepackage{amsmath,bm}
\usepackage{graphicx,epsfig,braket,amssymb,amsfonts,natbib}
\usepackage[caption=false]{subfig}
\DeclareGraphicsExtensions{.pdf,.png,.jpg}

\renewcommand{\v} [1]{{\bf #1}}
\newcommand{\ba}{\begin{eqnarray}}
\newcommand{\ea}{\end{eqnarray}}
\newcommand{\nn}{\nonumber \\}
\newcommand{\bpm}{\begin{pmatrix}}
\newcommand{\epm}{\end{pmatrix}}

\newcommand{\vep}{\varepsilon}
\newcommand{\lam}{\lambda}
\newcommand{\al}{\alpha}
\newcommand{\be}{\beta}
\newcommand{\ga}{\gamma}
\newcommand{\de}{\delta}
\newcommand{\sig}{\sigma}

\newcommand{\la}{\langle}
\newcommand{\ra}{\rangle}
\newcommand{\pd}{\partial}
\graphicspath{{figures/}}% Put all figures in this directory. Avoids clutter.

\begin{document}

\title{Thermal Hall Effect of Spins in a Paramagnet}

\author{Hyunyong Lee}
\affiliation{Department of Physics, Sungkyunkwan University, Suwon
  440-746, Korea}
\author{Jung Hoon Han}
\email[Electronic address:$~~$]{hanjh@skku.edu}
\affiliation{Department of Physics, Sungkyunkwan University, Suwon 440-746, Korea}
\author{Patrick A. Lee}
\email[Electronic address:$~~$]{palee@mit.edu}
\affiliation{Department of Physics, Massachusetts Institute of
Technology, Cambridge, Massachusetts 02139, USA}

\begin{abstract}
  Theory of Hall transport of spins in a correlated paramagnetic phase
  is developed. By identifying the thermal Hall current operator in the
  spin language, which turns out to equal the spin chirality in the
  pure Heisenberg model, various response functions can be derived
  straightforwardly. Subsequent reduction to the Schwinger boson
  representation of spins allows a convenient calculation of thermal
  and spin Hall coefficients in the paramagnetic regime using
  self-consistent mean-field theory. Comparison is made to results
  from the Holstein-Primakoff reduction of spin operators appropriate
  for ordered phases. 
\end{abstract}
 \maketitle

\section{introduction}

The Hall effect of electrons has evolved from a useful tool for measuring the carrier density of a material to a powerful diagnostic of the topological structure of the underlying electronic band, reflecting the Berry curvature distribution throughout the Brillouin zone\,\cite{nagaosa-review,niu-review}. Hall effect of charge current often implies the Hall effect for the energy, or of thermal transport, as the motion of electrons necessarily involves the transport of energy as well. 

Exciting recent developments have been the realization that this notion of topology-driven Hall effect can be extended to neutral objects of zero electrical charge. Phonon Hall effect, in which a transverse heat transport is mediated by phonons in response to thermal gradient, has been observed\,\cite{phonon-Hall-effect}. Magnons - quantized small fluctuations of an ordered magnet - can in principle exhibit similar Hall transport driven by thermal gradient, as first predicted theoretically by Katsura, Nagaosa, and Lee\,\cite{katsura2010} and confirmed experimentally in an insulating pyrochlore magnet Lu$_2$V$_2$O$_7$ by the Tokura group\,\cite{tokura2010}. 
Formulation of the magnon Hall effect was perfected by Murakami and collaborators in a series of papers\,\cite{murakami2011a,murakami2011b,murakami2014} after correcting for the missing, magnetization current term in the original derivation of Ref. \onlinecite{katsura2010}. A striking parallel of the topology of the magnon band structure to that of electronic bands responsible for quantized Hall effect was emphasized in several recent papers\,\cite{topological-magnon-paper2013,mook2014}. 

With a solid theoretical foundation and an experimental demonstration to back it up, the thermal Hall effect has become a powerful probe of the topological nature of magnon excitations in an ordered magnet. While the magnon Hall effect is easily interpreted as a natural consequence of momentum-space topology of the magnon band, a complementary real-space picture suggests that it is also a probe of a particular type of spin correlations, known as the spin chirality, of quantum insulating magnetic
systems\,\cite{tokura2010,katsura2010}. Spin chirality, expressed as the triple product of three neighboring spin operators $\v S_i \cdot \v S_j \times \v S_k$ for sites $i,j,k$ forming the smallest triangle in the lattice, has taken on the significance of an important new order parameter of a quantum spin system since its invention in the late 80's\,\cite{laughlin1987,wen1989}. An appealing possibility entertained ever since its inception is that of a quantum-disordered magnet with zero average local magnetization $\langle \v S_i \rangle =0$ yet with a finite spin chirality, $\langle \v S_i \cdot \v S_j \times \v S_k \rangle \neq 0$. Such a state breaks time-reversal symmetry and parity, opening the door for finite Hall-type transport in its ground state. A well-deserving question in this regard is whether the magnon Hall effect has a natural extension to the disordered phase, in which the notion of magnon may break down but not that of the spin chirality order. In other words, is the establishment of spin chirality (without the magnetic long-range order) a sufficient condition to give rise to thermal Hall effect in an insulating magnet? 

We will argue in this paper that there is no physical principle preventing the persistence of Hall-type transport into the paramagnetic phases of spin once the time reversal symmetry is broken by
the magnetic field. Thermal Hall measurement was successfully carried
out both below and {\it above the ferromagnetic transition temperature} in a different material by the Ong
group. \,\cite{hirschberger} Recently the same group shows the presence of thermal Hall effect in the frustrated (i.e. disordered) quantum pyrochlore material Tb$_2$Ti$_2$O$_7$.\,\cite{hirschberger2015} Stimulated  by their observations,
we go beyond the existing magnon description of the thermal Hall
effect\,\cite{katsura2010,tokura2010,murakami2011a,murakami2011b,murakami2014,topological-magnon-paper2013,mook2014}
and formulate the phenomenon using the spin language entirely. It is
then applied to discuss Hall effects of spin both in the paramgnetic
as well as the ferromagnetic regime. Essentially, the idea is to
develop the linear response formalisms within the spin language as
much as possible. Only in the final stage of the computation of the response
function is the particular representation of the spin operator relevant. For
instance the Hall effect in the ordered phase is appropriately captured by
the Holstein-Primakoff (HP) mapping of spins,  as had been done in the
past,\,\cite{katsura2010} while the possible paramagnetic Hall effect
is best discussed in the Schwinger boson (SB)
language.\,\cite{auerbach88, auerbach94} Both thermal and spin Hall
effects can be consistently described in this new formalism. 

In Sec. II we describe the new linear response formalism for
calculating thermal Hall conductivity entirely in the spin language,
followed in Sec. III by an explicit calculation of the thermal and
(related) spin Hall conductivities using the two well-known
approximate methods: Holstein-Primakoff and Schwinger boson
methods. Discussions and future prospects are given in Sec. IV.

\section{spin linear response theory}

\begin{figure}[!Ht]
  \includegraphics[width=0.25\textwidth]{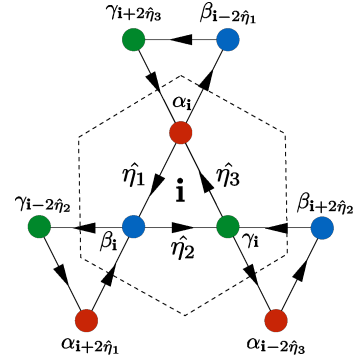}
  \caption{(Color online) Schematic figure of the Kagome lattice. Arrows
    indicate the sign convention $D_{i;j} = +D$ for $i\rightarrow j$. Unit vectors are chosen
    $\hat{\eta}_1=-(1,\sqrt{3})/2$, $\hat{\eta}_2=(1,0)$,
    $\hat{\eta}_3=(-1,\sqrt{3})/2$ with lattice constant
    $a=1$. Each upward triangle $\v i$ has three sublattice sites $\al_{\v i} ,\be_{\v i}, \ga_{\v i}$. }
  \label{fig:Kagome}
\end{figure}

To present the method of approach in a concrete
background we choose the Heisenberg spin model on a Kagome lattice,
written as a sum of site Hamiltonians $H = \sum_i H_i$, where each $H_i$ is

\ba
H_i \!=\!  \frac{1}{2} \sum_{j\in i} \Bigl( -J \v S_i \cdot \v S_j \!+\! D_{i;j} \v S_i
\times \v S_j \cdot \hat{z} \Bigr)  \!-\! B \v S_i \cdot \hat{b}.
\label{eq:h_local}
\ea
The symbol $j \in i$ indicates four immediate neighbors of each
site $i$. The orientation of the external field is fixed: $\hat{b}=+\hat{z}$. Nearest-neighbor exchange interaction of strength $J$ is assumed,
with the convention for the sign of the Dzyaloshinskii-Moriya (DM)
interaction $D_{i;j} = D=- D_{j;i} $ as outlined in
Fig.\,\ref{fig:Kagome}. Although all formal derivations of
spin linear response functions apply for either signs of $J$, for concreteness
we will assume ferromagnetic exchange $J>0$.

Two continuity equations are derived,

\ba \dot{S^z_i} + \sum_{j\in i} J^S_{i; j}  = 0 ,
~~\dot{H_i} + \sum_{j\in i} J^E_{i; j} = 0 ,
\label{eq:continuity}
\ea
tied to total $z$-spin and energy conservations, respectively. The bond current operators are

\begin{eqnarray}
  J^S_{i; j} &=& -i \frac{J'}{2} e^{i\phi_{i;j}} S_i^ + S_j^- + h.c.,\nn
  J^{E}_{i; j} &=& - B J_{i ; j}^S
  - \frac{1}{2} \sum_{k \in j } \Big( J S_{k}^z J_{i;j}^S
  + J S_i^z J_{j;k}^S
  +[J_{i;j}^S,\, J_{j;k}^S]\Big) \nn
  &&  +\frac{1}{2} \sum_{k \in i } \Big( J S_{k}^z J_{j;i}^S
  + J S_j^z J_{i;k}^S
  +[J_{j;i}^S,\, J_{i;k}^S]\Big).
  \label{eq:e_current_1}
\end{eqnarray}
The spin current $J^S_{i;j}$ for the $z$-component is expressed in terms of
$S_i^{\pm}  = S_i^x \pm i S_i^y$, $J' =\sqrt{J^2 + D^2}$, and $\tan \phi_{i;j} =
D_{i;j}/J$. While the spin current operator above is well known,
the energy current $J^E_{i;j}$ is new. In the Heisenberg limit ($D=0$) the energy current is directly related to the spin chirality,

\ba J^E_{i;j} = J^2 \sum_{k\in j}  \v S_i \cdot (\v S_j \times \v S_k ) ~~~ (D=0) . \label{eq:thermal-current-is-chirality} \ea
Linear response theory for the average of spin and energy current operators can be developed
now.

Coupling of the energy density
$H_i$ to the pseudo-gravitational potential $\psi_i$ is an
effective way to derive the thermal response function.\,
\cite{luttinger1964,murakami2011a,murakami2011b,murakami2014}
In brief, the total Hamiltonian including the gravitational coupling
$H =  \sum_i [ 1 + \psi_i e^{st} ] H_i$
%  \label{eq:h_t_local}
%
leads to the modification of the density matrix $\rho(t) = \rho_0 +
\de \rho\, e^{st}$,\cite{luttinger1964}

\begin{eqnarray}
  \de \rho
  &=& -\frac{\rho_0}{\hbar} \int_0^{\infty} dt' e^{-st'}
  \int_0^{\be} d\be' \sum_{\la i,j\ra} (\psi_j \!-\! \psi_i)
  J_{i; j}^E (-t'-i\be') \nn
  &\simeq& -\frac{\rho_0}{\hbar} \int_0^{\infty} dt' e^{-st'}
  \int_0^{\be} d\be' \sum_{\triangle_{\v i}}
  (\bm \nabla \psi) \cdot \v j_0^E(\v i;-t'-i\be') .\nn
  \label{eq:f}
\end{eqnarray}
The first line involves the sum over all nearest neighbors $\langle
i,j\rangle$ of the Kagome lattice, which in the second line is
re-organized as a sum over each upward-pointing triangle
$\triangle_{\v i}$. Assuming smoothly varying field allows one to
replace $\psi_j -\psi_i$ by its gradient. The ensuing current vector
$\v j_0^E (\v i )$ per triangle $\v i$ is a sum, 

\ba
  j_{0x}^E (\v i) &=&
  J_{\be_{\v i} ; \ga_{\v i}}^E
  + J_{\ga_{\v i \!-\! 2\hat{\eta}_2} ;  \be_{\v i}}^E\nn
&&
  + \frac{1}{2} \Big( J_{\be_{\v i} ; \al_{\v i}}^E
  \!+\! J_{\al_{\v i} ; \be_{\v i \! -\! 2\hat{\eta}_1}}^E
  \!+\! J_{\al_{\v i} ; \ga_{\v i}}^E
  \!+\! J_{\ga_{\v i} ; \al_{\v i \!-\! 2\hat{\eta}_3}}^E \Big), \nn
  j_{0y}^E (\v i) &=&\frac{\sqrt{3}}{2}\Big(
  J_{\be_{\v i} ; \al_{\v i}}^E
  \!+\! J_{\ga_{\v i} ; \al_{\v i}}^E
  \!+\! J_{\al_{\v i} ; \ga_{\v i \!+\! 2\hat{\eta}_3} }^E
  \!+\! J_{\al_{\v i} ; \be_{\v i \!-\! 2\hat{\eta}_1} }^E \Big),
  \label{eq:current_operator}
\end{eqnarray}
where all the subscript symbols are as defined in Fig. \ref{fig:Kagome}.

\begin{figure}[!Ht]
  \includegraphics[width=0.5\textwidth]{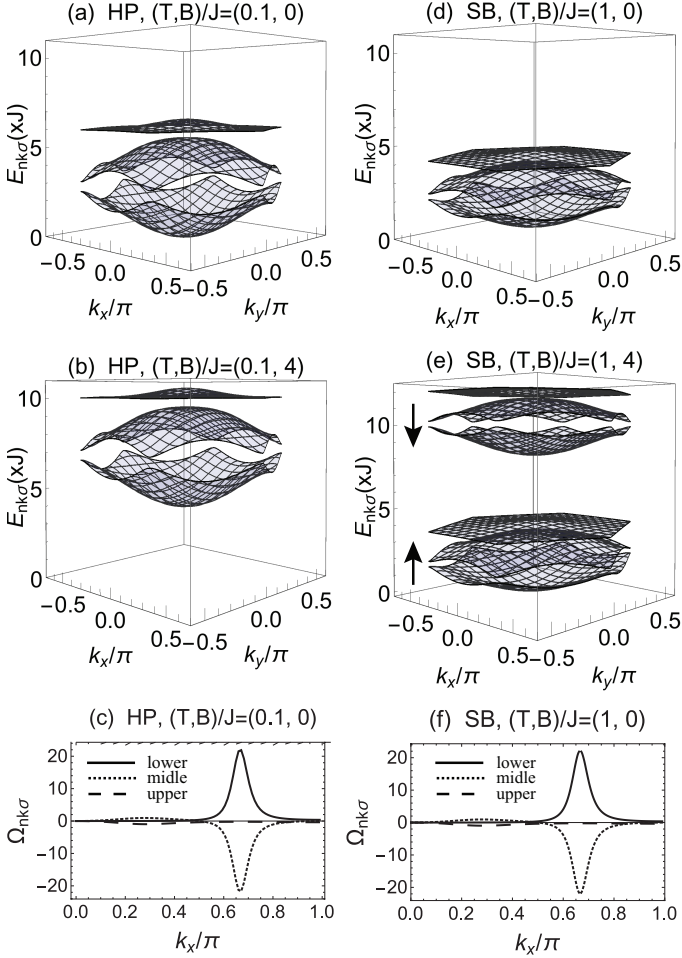}
  \caption{Magnon band structure from HP
    theory\,$(D/J=0.125)$ at (a) zero and (b) large Zeeman fields. SB
    band structure for $D/J=0.125$ at (d) zero and (e) large Zeeman
    fields. Berry curvatures are worked out at $k_y =0$ for (c) magnon
    bands and (f) SB bands of up spins. Hall responses are a
    consequence of population of low-lying magnon or Schwinger boson
    bands, multiplied by their respective Berry curvature densities in
    momentum space. Down-spin SB bands have exactly opposite Berry
    curvatures at $B=0$ and a similar one throughout $B>0$.} 
\label{fig:band}
\end{figure}

As noted long ago by Luttinger,\,\cite{luttinger1964} the psuedo-gravitational
field entering in the total Hamiltonian alters more than the density
matrix, as is often the assumption in linear response theory. Working
through the continuity equation for the modified local Hamiltonian
$(1\!+\!\psi_i ) H_i$ gives the new bond energy current operator

\ba
(1 \!+\! \psi_i \!+\! \psi_j ) J_{i;j}^E \simeq [1 + 2( \v r_{\v
  i} \cdot {\bm \nabla} \psi ) ] J_{i;j}^E . 
\ea
The failure of the local Hamiltonians to commute with each other, $[H_i ,
H_j ] \neq 0$, is the source of the modification. Such modification does not occur for instance in the case of electric current, since density operators (which couple to electric potential) commute at different sites.
The relation $\psi_i = \v r_{\v i}\cdot ({\bm \nabla}
\psi)$ for the uniform potential gradient is assumed. The physical energy
current operator is therefore the sum,

\begin{eqnarray}
  \v j^E(\v i)
  &=& \v j_{0}^E(\v i) + \v j_{1}^E(\v i), \nn
  \v j_{1}^E (\v i)
  &=& 2 \v j_{0}^E(\v i)( \v r_{\v i} \cdot {\bm \nabla} \psi).
\end{eqnarray}
Average of the energy current operator in response to the
pseudo-gravitational field accordingly contains two contributions,

\begin{eqnarray}
  \la j_{a}^E \ra &=&{\rm Tr}[\de \rho\,j_{0a}^E ]
  \!+\! {\rm Tr} [\rho_0\,j_{1a}^E ]
 = ( \sig_{0ab}^E  \!+\!
  \sig_{1ab}^E) (- {\bm \nabla}_{b}\psi).\nn
\end{eqnarray}
Spatial average $(1/N_t)\sum_{\triangle_{\v i}} \v j^E_0 (\v i ) \equiv \v j^E$,
$N_t$=number of up triangles, is taken. Formal expressions of these
coefficients are well-known and reproduced,

\begin{eqnarray}
&&  \sig_{0 a b}^E = \frac{i}{N_t}
  \sum_{n,m} \frac{e^{-\be \vep_m} \!-\! e^{-\be \vep_n}}{\vep_m \!-\! \vep_n}
  \frac{\la n |j_{0b}^E| m \ra \la m | j_{0a}^E | n \ra}{\vep_n
    \!-\!\vep_m \!-\! i s} , \nn
&& \sig_{1ab}^E = {\rm Tr}\left( \rho_0
  \left[ \frac{ \partial j_{0a}^E(\v q) }{ \partial q_b } \right]_{\v q=0} \right) ,
\label{eq:sigma_ab}
\end{eqnarray}
where complete sets of many-body states are $|m\rangle$ and
$|n\rangle$ and ${\v j}_0^E (\v q) = (1/N_t) \sum_{\triangle_{\v i}} {\v j}_0^E (\v i)
e^{-i \v q \cdot \v r_{\v i}}$.

This completes the derivation of thermal response functions in the
spin language. To evaluate them, however, is hard without a full
knowledge of all many-body eigenstates for the spin Hamiltonian. Below
we propose a scheme in which evaluation of $\sigma^E_{ab}=\sig_{0ab}^E
\!+\! \sig_{1ab}^E$ can be performed straightforwardly at the non-interacting level.
\\

\section{ Holstein-Primakoff and Schwinger boson linear response theory}

Evaluation of the response coefficients can
be done in the Schwinger boson mean-field theory (SBMFT) in which spin
is expressed by a pair of bosons $(b_{i\uparrow} , b_{i\downarrow})$ as $\v S_i =
\frac{1}{2} \sum_{\al,\be=\uparrow,\downarrow} b_{i\al}^{\dag} {\bm \sig}_{\al\be}
b_{i\be}$. Decoupling in terms of the bond operator
$\hat{\chi}_{i;j}^{\sig} = b_{i\sig}^{\dag} b_{j\sig}$ gives the mean-field Hamiltonian,

\begin{eqnarray}
  H^{\rm SB} &=& \sum_{i,\sig} (\lam - \sig B) b_{i\sig}^{\dag} b_{i\sig}
  - \sum_{\la i,j \ra, \sig} \left( t^{\sig}_{i;j}
  b_{i\sig}^{\dag}b_{j\sig} + h.c. \right),\nn
  t^{\sig}_{i;j} &=& J \la \hat{\chi}_{j;i}^{\sig} \ra
  + J' e^{-i\sig\phi_{i;j}} \la \hat{\chi}_{j;i}^{-\sig}\ra .
  \label{eq:h_schwinger}
\end{eqnarray}
The Lagrange multipler $\lambda$ is introduced to keep the average boson
number constant at $2S=1$. The Zeeman field and the effective flux from DM
interaction act oppositely for the two bosons. The energy
current operator in Eq.\,\eqref{eq:e_current_1} allows a lengthy
re-writing in terms of bond operators

\begin{widetext}
\begin{eqnarray}
  J_{i;j}^E &=&
  -{1\over2} B (J \!+\! iD_{i;j} ) \sum_{\sig}
  \hat{\chi}_{i;j}^{- \sig} \hat{\chi}_{j;i}^{\sig}
  +\frac{1}{16i} \sum_{k \in j}\Bigg\{
  J^2 (\hat{\chi}_{i;j} \hat{\chi}_{j;k} \hat{\chi}_{k;i} - h.c.)
  + D_{i;j} D_{j;k} \sum_{\sig}
  \left( \hat{\chi}_{i;j}^{-\sig} \hat{\chi}_{j;k}^{-\sig}
    \hat{\chi}_{k;i}^{\sig} - h.c. \right) \nn
  && + i J \sum_{\sig} \sig \left(
    D_{i;j} \hat{\chi}_{i;j}^{-\sig} \hat{\chi}_{j;k} \hat{\chi}_{k;i}^{\sig}
    + D_{j;k} \hat{\chi}_{k;j}^{-\sig} \hat{\chi}_{j;i} \hat{\chi}_{i;k}^{\sig}
    + h.c. \right) \Bigg\}
  - (i \leftrightarrow j), \label{eq:J-E-chi}
\end{eqnarray}
\end{widetext}
where $\hat{\chi}_{i;j} = \sum_{\sig} \hat{\chi}_{i;j}^{\sig}$, and
$(i\leftrightarrow j)$ denotes the exchange for all the terms shown in Eq. (\ref{eq:J-E-chi}).

Due to the enormous complexity of the current operator in the
Schwinger boson representation (or in the spin representation for that
matter), calculating the correlation function for it appears daunting if not impossible.
However, one observes that each triple product
of bond operators in the above expression contains exactly two terms that can
be replaced by the mean-field average $\langle
\hat{\chi}_{i;j}^\sigma\rangle$ (because they span the nearest
neighbours in the Kagome lattice), and only one that contains boson
hopping across second neighbors (not captured by the
mean-field parameterization). After such mean-field reduction $J^E_{i;j}$ becomes a
bilinear in the Schwinger boson operator\,[see Appendix A]. In the uniform
case, $\langle \hat{\chi}_{i;j}^\sigma \rangle = \chi_\sigma$, we have proven that the corresponding mean-field vector current operator
$\v j_0^E (\v i)$, averaged over all triangles $\v j^E_0 = (1/N_t )
\sum_{\triangle_{\v i}} \v j^E_0 (\v i)$, is equal to a simple and familiar
expression\,[see Appendix A]

\begin{eqnarray}
  \v j^E_0 &=&  {1\over 2} \sum_{\v k, \sig}\Psi^\dag_{\v k \sigma}  \left(
    H^{\rm SB}_{\v k\sig} \frac{\pd H^{\rm SB}_{\v k\sig}}{\pd \v k}
    +\frac{\pd H^{\rm SB}_{\v k\sig}}{\pd \v k}  H^{\rm SB}_{\v k\sig} \right) \Psi_{\v
    k \sigma} .
\end{eqnarray}
We denote the three corners of the upward triangle $\v i$ as $\alpha_{\v i},
\beta_{\v i}, \gamma_{\v i}$, respectively (Fig.\,\ref{fig:Kagome}),
and their Fourier counterparts as $\Psi^T_{\v k \sigma} = ( \al_{\v k
  \sig} ~ \be_{\v k \sig} ~\ga_{\v k \sig} )$. Mean-field SB
Hamiltonian in Eq.  (\ref{eq:h_schwinger}) for uniform parameters
becomes in momentum space $H^{\rm SB} =\sum_{\v k, \sig}\Psi^\dag_{\v
  k \sigma}  H^{\rm SB}_{\v k\sig} \Psi_{\v k \sigma}$,

\begin{eqnarray}
&& H^{\rm SB}_{\v k\sig} \! = \! (\lambda \! -\! \sig B) I_3 \!+\!
  \begin{pmatrix}
    0 & t_{\sig} \cos k_1
    & t_{\sig}^*\cos k_3\\
   t_{\sig}^*\cos k_1  & 0
    &t_{\sig} \cos k_2 \\
     t_{\sig}\cos k_3
    &t_{\sig}^*\cos k_2 & 0
  \end{pmatrix}, \nn
\label{eq:mf_h}
\end{eqnarray}
with effective hopping parameters $t_{\sig} = J \chi - i \sig D
\chi_{-\sig}$, $\chi = \sum_\sigma \chi_\sigma$, $k_x = \v k \cdot
\hat{\eta}_x$ and $\hat{\eta}_x$ are the three orientation unit
vectors defined in Fig.\,\ref{fig:Kagome}. We note that for each spin
$\sigma$, both the current operator $\v j^E_{\v k\sigma}$ and the
Hamiltonian $H_{\v k \sigma}$ have identical forms as those already
examined for magnon thermal Hall problem on the Kagome
lattice.\,\cite{katsura2010, murakami2014} Thus, known thermal Hall
formulas derived previously can be applied here directly,
for evaluation {\it in the paramagnetic regime}.

%\onecolumngrid
\begin{figure*}[!Ht]
\includegraphics[width=1.\textwidth]{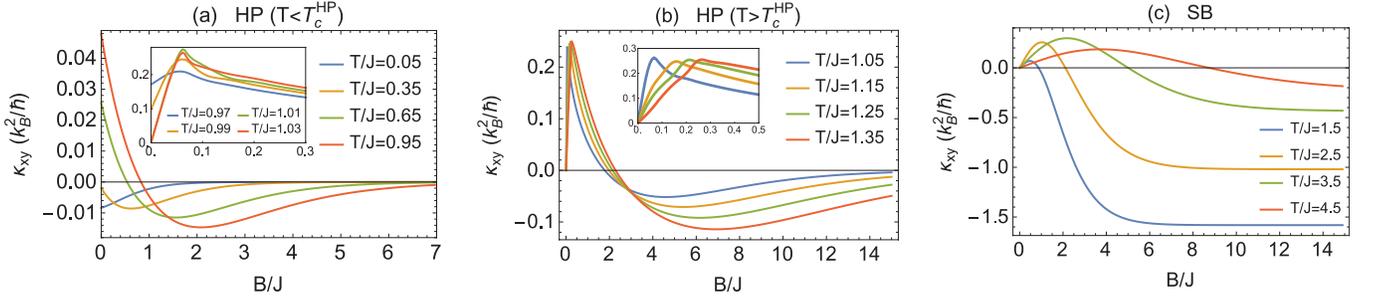}
\caption{(Color online) Low-temperature thermal Hall conductivity
  based on self-consistent HP theory ($D/J=0.125$) for (a) $T < T^{\rm
    HP}_c$ and (b) $T > T^{\rm HP}_c$. Zero-field ferromagnetic
  transition occurs at  $T_c^{\rm HP}
  / J = 1$. Inset to (a) highlights the sensitive dependence of $\kappa^{\rm HP}_{xy}$ on temperature around $T=T^{\rm HP}_c$ due to small values of self-consistent magnetization $S$ and the consequent collapse of magnon bands, leading to a large enhancement of the Bose factor in Eq. (\ref{eq:kappa_xy}) over a small temperature change. Inset to (b) emphasizes the linear rise of $\kappa^{\rm HP}_{xy}$ with magnetic field for $T>T^{\rm HP}_c$. (c)
  High-temperature thermal Hall conductivity based on SBMFT
  ($D/J=0.15$) for $T > T^{\rm SB}_c (\approx 0.5 J)$.}
\label{fig:kappa}
\end{figure*}

\begin{figure*}[!Ht]
\includegraphics[width=1.\textwidth]{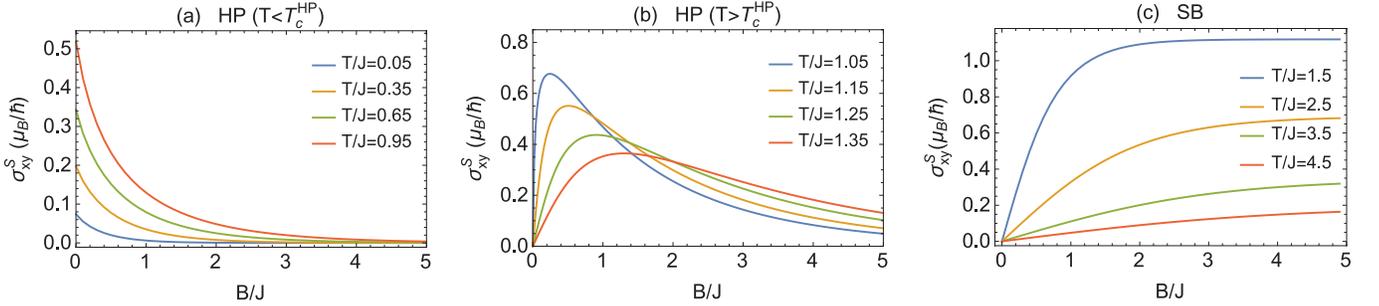}
\caption{(Color online) Spin Hall conductivity $\sigma_{xy}^S$ based
  on HP theory ($D/J=0.125$) for (a) $T < T^{\rm HP}_c$ and (b)  $T >
  T^{\rm HP}_c$. (c) High-temperature spin Hall conductivity based
  on SBMFT ($D/J=0.15$) for $T > T^{\rm SB}_c$.}
\label{fig:spin-Hall}
\end{figure*}

The thermal Hall conductivity within the SB theory reads

\begin{eqnarray}
  \kappa^{\rm SB}_{xy} &=& - \frac{k_B^2 T}{\hbar N_t}
  \sum_{\v k, n, \sig} \left[
    c_2\left( E^{\rm SB}_{n\v k \sig} \right) - \frac{\pi^2}{3}
  \right] \Omega^{\rm SB}_{n\v k \sig} . \label{eq:kappa_xy}
\end{eqnarray}
Both the energy dispersions and Berry curvatures are to be obtained
from diagonalizing the Hamiltonian, Eq. (\ref{eq:mf_h}), $c_2(x) =
(1+x)\left( \ln \frac{1+x}{x} \right)^2 - \left(\ln x\right)^2 - 2
{\rm Li}_2(-x)$,\,\cite{murakami2014} and $\Omega^{\rm SB}_{n \v k
  \sig} = i\langle\partial_{k_x}  u_{n\v k \sig} | \partial_{k_y}
u_{n\v k \sig}\rangle + c.c.$ for the $n$-th band eigenstate $\left|
  u_{n\v k \sig} \right\ra$ of $H^{\rm SB}_{\v k \sig}$.

By comparison, HP substitution of spin operators in the spin
Hamiltonian (\ref{eq:h_local})  leads to the familiar magnon
Hamiltonian\,\cite{katsura2010}

\begin{eqnarray}
  H^{\rm HP} \!=\!  - S J' \sum_{\la i, j \ra} (e^{-i \phi_{i;j}}
  b_i^{\dag} b_j \!+\! h.c.) \!+\! \sum_i (B\!+\!4SJ) b_i^{\dag} b_i
  ,\nonumber
\label{eq:magnon}
\end{eqnarray}
where $S$ is the size of average magnetization, either by spontaneous
order or through external field. Different from earlier
work\,\cite{katsura2010} we invoke self-consistency relation 

\ba S(B,T)= {1\over 2} \tanh\left[{ B\!+\! 4JS(B,T) \over 2k_B T }\right] \nonumber \ea 
to work out $S$ at a given temperature and field strength. Spontaneous magnetization
$S\neq0$ occurs at $T_c^{\rm HP} = J$. Magnon thermal Hall formula $\kappa^{\rm HP}_{xy}$ is obtained the same way as in Eq. (\ref{eq:kappa_xy}) without the sum over the spin index $\sigma$.

Figure\,\ref{fig:band} shows representative band dispersions and Berry
curvature distributions over the first Brillouin zone for SB and HP
bosons, respectively. At $B=0$ both SB bands look nearly
identical to the magnon bands except for the non-zero band minimum
(SB bosons are not Goldstone bosons). The zero-field Berry curvatures
are also quite similar for SB and HP bosons, as shown in
Fig.\,\ref{fig:band}, but not
identical, because effective DM constant in the SB theory
is halved, $t_\sigma = J \chi - i\sigma \chi_{-\sigma} = \chi
(J-\sigma D/2)$, at $B=0$.

Figure \ref{fig:kappa} displays thermal Hall response coefficients
from HP and SB theories. Recall that $\kappa^{\rm HP}_{xy}$ is finite
even at zero field, due to the spontaneous flux generated by the DM
interaction.\,\cite{katsura2010} The $B=0$ value however changes sign
upon raising the temperature as shown in Fig. \ref{fig:kappa}(a),
because the higher magnon band has the opposite Berry curvature as
shown in Fig. \ref{fig:band}(c). On further increase of $T$ it goes
down to zero at $T=T_c^{\rm HP}$. There is also a sign reversal of the
Hall response at finite field, in qualitative agreement with the recent measurement reported by the Ong
group\,\cite{hirschberger}. At low temperature and low field the
lowest-lying magnon band dominates transport. For higher temperatures,
higher-energy band carrying opposite Berry flux (see Fig. 2(c)) has a
chance to contribute significantly. Strong Zeeman field creates a
large gap for all the bands, diminishing the thermal population
difference among the bands and increasing the relative contribution of
the higher band with significant Berry flux concentration. The
Schwinger boson Hall transport, shown in Fig. \ref{fig:kappa}(c), is
already at quite high a temperature and continues the trend seen in
the high-temperature magnon calculation, i.e. a positive peak at low
field followed by a long negative tail in the high-field
region. Together, we are assured that thermal Hall transport is a
sensitive probe of the Berry flux distribution as well as the band
structure of the underlying elementary excitations in an insulating
paramagnet.

The spin Hall response can be worked out in much
the same way by replacing the spin current operator in
Eq. (\ref{eq:e_current_1}) with its mean-field version\,[see Appendix B]. The source
term for spin current, $-\sum_i h_i S^z_i$, does not modify the spin
continuity equation  since $[S_i^z , S_j^z ] =0$. Mean-field spin
current operator 

\ba \v j^S = \sum_{\v k, \sigma} \sigma \Psi^\dag_{\v
  k\sigma} { \partial H_{\v k \sigma} \over  \partial \v k } \Psi_{\v k
  \sigma} \nonumber \ea 
results in the spin Hall conductivity   

\ba 
\sigma_{xy}^{S, \rm SB}= {\mu_B \over \hbar N_t } \sum_{\v k, n,
  \sigma} n_B(E_{n \v k \sig}) \Omega^{\rm SB}_{\v k, n, \sigma}, 
\ea 
where $n_B$ is the Bose occupation function.  Spin Hall coefficients
for both HP and SB boson theories are worked out in
Fig. \ref{fig:spin-Hall}.

\section{ Discussion}

Theories of thermal and spin Hall
effects for spin systems are developed in the general language of spin
operators. Ways to consistently obtain response functions in the
correlated disordered phase are developed, employing Schwinger boson
approach. The Holstein-Primakoff reduction is shown to reproduce the
existing theories. Most interestingly, Eq. (\ref{eq:thermal-current-is-chirality})
unambiguously points out that thermal Hall response is a direct
measure of the inherent spin chirality in the underlying system, along
with other spectroscopic probes of spin chirality recently
proposed\,\cite{wingho2011, patrick2013}. As our derivations in Sec. II do not assume a particular lattice geometry, the formalisms developed in this paper will be applicable to spin models defined on any lattice geometry in both two and three dimensions. 

Regarding the actual computation of the thermal and spin Hall response functions we have employed self-consistent Holstein-Primakoff and Schwinger boson methods in this paper. Other means of computing the thermal Hall coefficients in the spin system, such as exact diagonalization, can be an alternative to the
methods presented in this paper. There are shortcomings in the
so-called ``exact methods" due to the severe size limitations in the
diagonalization and the difficulty of extrapolating the computation to
large system size. The abundance of low-energy states that are crucial
to efficient thermal transport may be difficult to capture in the
exact diagonalization on small system size. On the other hand, the
mean-field nature in the Schwinger boson approach calls for
improvements in regard to effects of fluctuations~\cite{auerbach88,sachdev92,auerbach94}. 
In particular the phase fluctuation in the mean-field order parameter $t^\sigma_{i;j}$ may remain gapless and severely disrupt the mean-field analysis unless well-known mass-generating mechanisms (such as Anderson-Higgs or Chern-Simons) play a role. We plan to complement the present work, focused on the formulation of spin thermal transport and its evaluation in the simplest possible manner,
in several directions with the forthcoming publication with emphasis on the importance of gauge fluctuations in the Schwinger boson formalism. 

\acknowledgments{J. H. H. is supported by the NRF grant
(No. 2013R1A2A1A01006430) and wishes to acknowledge the hospitality
of the condensed matter theory group at MIT and Boston College where
this work was carried out. P.A.L. acknowledges support by the DOE
grant DE-FG01-03-ER46076. We are grateful to Robin Chisnell, Young S. Lee, and Phuan Ong for many inspiring discussions.}  

\bibliographystyle{apsrev}
\bibliography{reference}

\onecolumngrid
\appendix

\section{Energy current operator in the Schwinger boson mean field
  theory} 

The bond energy current operator appearing in the continuity equation
$\dot{H_i} + \sum_j J^E_{i; j} = 0$ was written in terms of Schwinger
boson operator in the following way,

\begin{eqnarray}
%   J^{E}_{i; j} 
%   &=& i\frac{B(J+iD_{i;j})}{2} S_i^+ S_j^- + h.c\nn
%   &-& \frac{1}{2} \sum_{k \in j } \Big( J S_{k}^z J_{i;j}^S
%   + J S_i^z J_{j;k}^S
%   +[J_{i;j}^S,\, J_{j;k}^S]\Big)
%   +\frac{1}{2} \sum_{k \in i } \Big( J S_{k}^z J_{j;i}^S
%   + J S_j^z J_{i;k}^S
%   +[J_{j;i}^S,\, J_{i;k}^S]\Big).
  J_{i;j}^E 
  &=& -{1\over 2} B (J + iD_{i;j} ) \sum_{\sig}
  \hat{\chi}_{i;j}^{- \sig} \hat{\chi}_{j;i}^{\sig}\nn
  &+&\frac{1}{16i} \sum_{k \in j}\Bigg\{
  J^2 \left[ \hat{\chi}_{i;j} \hat{\chi}_{j;k}\hat{\chi}_{k;i} - h.c. \right]
 + D_{i;j} D_{j;k} \sum_{\sig}
  \left[ \hat{\chi}_{i;j}^{-\sig} \hat{\chi}_{j;k}^{-\sig}
    \hat{\chi}_{k;i}^{\sig} - h.c. \right]\nn
  &+& i J \sum_{\sig} \sig \left(
    D_{i;j} \hat{\chi}_{i;j}^{-\sig} \hat{\chi}_{j;k}\hat{\chi}_{k;i}^{\sig}
    + D_{j;k} \hat{\chi}_{k;j}^{-\sig} \hat{\chi}_{j;i}
    \hat{\chi}_{i;k}^{\sig}
    + h.c. \right) \Bigg\}
  - (i \leftrightarrow j).
\end{eqnarray}
This expression has six boson operators multiplied together and it
is impractical to carry out linear response calculations for it. On
implementing the mean field substitution for the nearest-neighbor bond
operators $\la \hat{\chi}_{i;j}^{\sig}\ra = \la b_{i\sig}^{\dag}
b_{j\sig} \ra \equiv \chi_{\sig}$ or $\chi_{\sig}^*$ following the
same convention as for DM interaction depicted in Fig.\, 1 of the main
text, we obtain the mean-field energy current operator 

\begin{eqnarray}
  J_{i;j}^E 
  &\stackrel{\rm SBMF}{\longrightarrow}&
  -{1\over 4} B (J + iD_{i;j} ) \sum_{\sig}
  \left[ \la \hat{\chi}_{i;j}^{- \sig} \ra \hat{\chi}_{j;i}^{\sig} 
    + \hat{\chi}_{i;j}^{- \sig} \la \hat{\chi}_{j;i}^{\sig} \ra \right]\nn
  &+&\frac{1}{16i} \sum_{k \in j}\Bigg\{
  J^2 \left[ \la \hat{\chi}_{i;j} \ra  \la \hat{\chi}_{j;k} \ra 
  \hat{\chi}_{k;i} - h.c. \right]
  + D_{i;j} D_{j;k} \sum_{\sig}
  \left( \la \hat{\chi}_{i;j}^{-\sig} \ra \la \hat{\chi}_{j;k}^{-\sig} \ra
    \hat{\chi}_{k;i}^{\sig} - h.c. \right)\nn
  &+& i J \sum_{\sig} \sig \left(
    D_{i;j} \la \hat{\chi}_{i;j}^{-\sig} \ra \la \hat{\chi}_{j;k} \ra 
    \hat{\chi}_{k;i}^{\sig}
    + D_{j;k} \la \hat{\chi}_{k;j}^{-\sig} \ra \la \hat{\chi}_{j;i}\ra 
    \hat{\chi}_{i;k}^{\sig}
    + h.c. \right) \Bigg\}
  - (i \leftrightarrow j) .
\end{eqnarray}
Only the bond operators connecting second-nearest neighbors remain as operators now. 
It is a boson bi-linear.  Here the MF
parameter substitution needs to be done carefully, because it could be
either $\chi_{\sig}$ or $\chi_{\sig}^*$ depending $i$ and $j$ as
explained before. Using the above expression and Eq. 5 of the main text (reproduced here)

\ba 
  j_{0x}^E (\v i) &=&
  J_{\be_{\v i} ; \ga_{\v i}}^E
  + J_{\ga_{\v i \!-\! 2\hat{\eta}_2} ;  \be_{\v i}}^E
  + \frac{1}{2} \Big( J_{\be_{\v i} ; \al_{\v i}}^E
  \!+\! J_{\al_{\v i} ; \be_{\v i \! -\! 2\hat{\eta}_1}}^E
  \!+\! J_{\al_{\v i} ; \ga_{\v i}}^E
  \!+\! J_{\ga_{\v i} ; \al_{\v i \!-\! 2\hat{\eta}_3}}^E \Big), \nn
  j_{0y}^E (\v i) &=&\frac{\sqrt{3}}{2}\Big(
  J_{\be_{\v i} ; \al_{\v i}}^E
  \!+\! J_{\ga_{\v i} ; \al_{\v i}}^E
  \!+\! J_{\al_{\v i} ; \ga_{\v i \!+\! 2\hat{\eta}_3} }^E
  \!+\! J_{\al_{\v i} ; \be_{i \!-\! 2\hat{\eta}_1} }^E \Big), \ea
one can convert the bond current to the vector current operator $j_{0x}^E(\v i)$ and $j_{0y}^E(\v i)$. Note that each bond current operator $J^E_{i;j}$ itself consists of dozen different terms as shown in Eq. (3) of the main article. Each vector current operator then consists of $\sim 10^2$ terms. Assignment of $\chi_\sigma$ or $\chi_\sigma^*$ for each average in the above equation (2) has to be carried out out term-by-term. Having completed such exercise, we finally arrive at the momentum-space expression for the current operator, 

\begin{eqnarray}
  j_{0\al}^E = \frac{1}{N_t} \sum_{\triangle_{\v i}} j_{0\al}^E(\v i)
  =\sum_{\v k, \sig} \Psi_{\v k \sig}^{\dag}
  \left( J^2 {\cal A}_{\al \v k} + JD \,{\cal B}_{\al \v k \sigma} + D^2 {\cal C}_{\al \v k \sigma}
  \right) \Psi_{\v k \sig}, 
  \label{eq:cur_op_spin_H}
\end{eqnarray}
where for the $x-$direction

{\tiny
\begin{eqnarray}
  &&{\cal A}_{x \v k} =
  \begin{pmatrix}
   \frac{|\chi|^2}{2}\left[ \sin 2k_1 + \sin 2k_3 \right]&
   -\frac{(\chi^*)^2}{4} \left[ 3 \sin(k_2 - k_3) + \sin(k_2 + k_3)\right]&
   \frac{\chi^2}{4} \left[ 3 \sin(k_2 - k_1) + \sin(k_2 +k_1)\right]\\
   -\frac{\chi^2}{4}\left[3 \sin(k_2 - k_3) + \sin(k_2 +k_3)\right]&
   \frac{|\chi|^2}{2} \left[ \sin 2k_1 - 2\sin 2 k_2 \right]& 
   \frac{(\chi^*)^2}{2} \sin(k_1 + k_3)\\
   \frac{(\chi^*)^2}{4} \left[ 3 \sin(k_2 - k_1) + \sin(k_2 +k_1)\right]&
   \frac{\chi^2 }{2} \sin(k_1 + k_3)&
   \frac{|\chi|^2}{2} \left[\sin 2k_3 - 2\sin 2k_2 \right]
 \end{pmatrix},\nn\nn
  &&{\cal B}_{x \v k \sigma} = \sig
  \begin{pmatrix}
   {\rm Im}[\chi\,\chi_{-\sig}^*] \left( \sin 2k_1 + \sin 2k_3 \right) &
  i (\chi\,\chi_{-\sig})^*\left[ \sin(k_3+k_2) - 3\sin(k_3-k_2)\right]&
  i \chi\,\chi_{-\sig} \left[\sin(k_2 + k_1) - 3\sin(k_2-k_1)\right]\\
  -i \chi\,\chi_{-\sig} \left[ \sin(k_3+k_2) - 3\sin(k_3-k_2)\right]&  
  {\rm Im}[\chi\,\chi_{-\sig}] \left(\sin 2k_1 - 2\sin 2 k_2\right)&
  -i (\chi\, \chi_{-\sig})^* \sin(k_1 + k_3)\\
  -i (\chi\,\chi_{-\sig})^* \left[\sin(k_2 + k_1) - 3\sin(k_2-k_1)\right]&
  i \chi\, \chi_{-\sig} \sin(k_1 + k_3)&
   {\rm Im}[\chi\,\chi_{-\sig}]\left( \sin 2k_3 - 2 \sin 2k_2 \right)
  \end{pmatrix},\nn\nn
  &&{\cal C}_{x \v k \sigma} =
  \begin{pmatrix}
   \frac{|\chi|^2}{2}\left[ \sin 2k_1 + \sin 2k_3 \right]&
   \frac{(\chi_{-\sig}^*)^2}{4} \left[ 3 \sin(k_2 - k_3) + \sin(k_2 +
     k_3)\right]& 
   -\frac{\chi_{-\sig}^2}{4} \left[ 3\sin(k_2 - k_1) + \sin(k_2 +k_1)\right]\\
   \frac{\chi_{-\sig}^2}{4}\left[3 \sin(k_2 - k_3) + \sin(k_2 +k_3)\right]&
   \frac{|\chi|^2}{2} \left[ \sin 2k_1 - 2\sin 2 k_2 \right]& 
   -\frac{(\chi_{-\sig}^*)^2}{2} \sin(k_1 + k_3)\\
   -\frac{(\chi_{-\sig}^*)^2}{4} \left[ 3 \sin(k_2 - k_1) + \sin(k_2
     +k_1)\right]& 
   -\frac{\chi_{-\sig}^2 }{2} \sin(k_1 + k_3)&
   \frac{|\chi|^2}{2} \left[\sin 2k_3 - 2\sin 2k_2 \right]
  \end{pmatrix},\nonumber
\end{eqnarray}
}
and for $y$-direction,

{\normalsize
\begin{eqnarray}
  &&{\cal A}_{y \v k} = \sqrt{3}
  \begin{pmatrix}
   \frac{|\chi|^2}{2}\left[ \sin 2k_1 - \sin 2k_3 \right]&
   -\frac{(\chi^*)^2}{4} \cos k_2 \sin k_3&
   \frac{\chi^2}{4} \cos k_2 \sin k_1\\
   -\frac{\chi^2}{2} \cos k_2 \sin k_3&
   \frac{|\chi|^2}{2} \sin 2k_1& 
   \frac{(\chi^*)^2}{2} \sin(k_1 - k_3)\\
   \frac{(\chi^*)^2}{4} \cos k_2 \sin k_1&
   \frac{\chi^2 }{2} \sin(k_1 - k_3)&
   - \frac{|\chi|^2}{2} \sin 2k_3
 \end{pmatrix},\nn\nn
  &&{\cal B}_{y \v k \sigma} = \sig \sqrt{3}
  \begin{pmatrix}
   {\rm Im}[\chi\,\chi_{-\sig}^*] \left[ \sin 2k_1 - \sin 2k_3 \right] &
  i (\chi\,\chi_{-\sig})^* \cos k_2 \sin k_3&
  i \chi\,\chi_{-\sig} \cos k_2 \sin k_1\\
  -i \chi\,\chi_{-\sig} \cos k_2 \sin k_3&  
  {\rm Im}[\chi\,\chi_{-\sig}] \sin 2k_1  &
  -i (\chi\, \chi_{-\sig})^* \sin(k_1 - k_3)\\
  -i (\chi\,\chi_{-\sig})^* \cos k_2 \sin k_1&
  i \chi\, \chi_{-\sig} \sin(k_1 - k_3)&
  {\rm Im}[\chi\,\chi_{-\sig}] \cos k_3 \sin k_3
 \end{pmatrix},\nn\nn
  &&{\cal C}_{y \v k \sigma} = \sqrt{3}
  \begin{pmatrix}
   \frac{|\chi|^2}{2}\left[ \sin 2k_1 - \sin 2k_3 \right]&
   \frac{(\chi_{-\sig}^*)^2}{4} \cos k_2 \sin k_3&
   -\frac{\chi_{-\sig}^2}{4} \cos k_2 \sin k_1\\
   \frac{\chi_{-\sig}^2}{4} \cos k_2 \sin k_3&
   \frac{|\chi|^2}{2} \sin 2k_1& 
   -\frac{(\chi_{-\sig}^*)^2}{2} \sin(k_1 - k_3)\\
   -\frac{(\chi_{-\sig}^*)^2}{4} \cos k_2 \sin k_1&
   -\frac{(\chi_{-\sig})^2}{2} \sin(k_1 - k_3)&
   -\frac{|\chi|^2}{2} \sin 2k_3
  \end{pmatrix}.\nonumber
\end{eqnarray}
}
\begin{figure*}[!Ht]
\includegraphics[width=.5\textwidth]{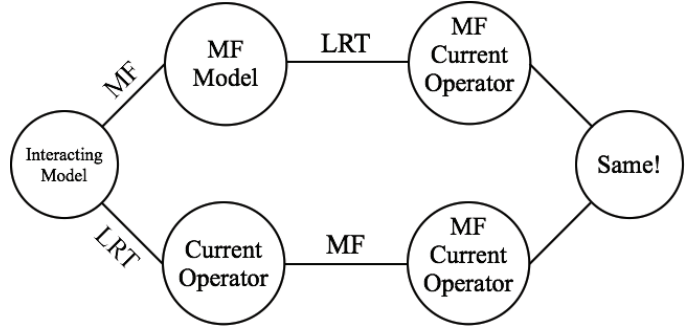}
\caption{Schematic figure demonstrating the equivalence we have
  provided in Appendix A. MF (mean field) and LRT (linear response
  theory) procedures can be interchanged, leading to the same, final
  linear response coefficients. }  
\label{fig:question}
\end{figure*}

Remarkably, the hopelessly lengthy expression found above is completely equal, term-by-term, to the following much simpler and intuitive expression

\begin{eqnarray}
  {\v j}_{0}^E = \frac{1}{2} \sum_{\v k , \sig}\Psi_{\v k \sig}^{\dag} \left(
    H_{\v k \sig} \frac{\pd H_{\v k \sig}}{\partial \v k}
    + \frac{\pd H_{\v k \sig}}{\partial \v k} H_{\v k \sig} \right) 
  \Psi_{\v k \sig}.
  \label{eq:cur_op_MF_H}
\end{eqnarray}
Here $H_{\v k \sigma}$ is the Schwinger boson mean-field Hamiltonian mapping of the original spin Hamiltonian. Reproducing Eq. (10) of the main text,

\begin{eqnarray}
  H^{\rm SB} &=& \sum_{i,\sig} (\lam - \sig B) b_{i\sig}^{\dag} b_{i\sig}
  - \sum_{\la i,j \ra, \sig} \left( t^{\sig}_{i;j}
  b_{i\sig}^{\dag}b_{j\sig} + h.c. \right),\nn
  t^{\sig}_{i;j} &=& J \la \hat{\chi}_{j;i}^{\sig} \ra
  + J' e^{-i\sig\phi_{i;j}} \la \hat{\chi}_{j;i}^{-\sig}\ra,
\end{eqnarray}
and making proper uniform-state ansatz $\langle \hat{\chi}_{i;j}^\sigma \rangle = \chi_\sigma ~ (\chi_\sigma^*)$ gives the momentum space Schwinger boson Hamiltonian [Eq. (13) of the main article] 

\begin{eqnarray}
&& H^{\rm SB}_{\v k\sig} \! = \! (\lambda \! -\! \sig B) I_3 \!+\!
  \begin{pmatrix}
    0 & t_{\sig} \cos k_1
    & t_{\sig}^*\cos k_3\\
   t_{\sig}^*\cos k_1  & 0
    &t_{\sig} \cos k_2 \\
     t_{\sig}\cos k_3
    &t_{\sig}^*\cos k_2 & 0
  \end{pmatrix} . 
\end{eqnarray}

Meaning of the complete equivalence we just obtained is given schematically in Fig. 1. One starts with an interacting spin model, derive the proper energy current operator from it, and then reduce it to its mean-field form (bottom path of the flow in Fig. 1). On the other hand, one can begin by writing down the mean-field Hamiltonian for the interacting spin model first, and derive the current operator from the mean-field, non-interacting Hamiltonian (top path of the flow). The results, as we demonstrate here, are identical. All the convenient machinery of linear response theory for non-interacting models can be brought to bear on the interacting problem now. 

\section{Spin current operator in Schwinger boson mean field theory}

As for the spin current operator, we can follow the same procedure developed for dealing with the energy current operator in the previous section. First one converts the bond spin current operator to the vector spin current according to Eq. (3) [Eq. (5) of main text], then take average over the whole lattice. In momentum space we get

\begin{eqnarray}
  J_{i;j}^S &=&  -\frac{i}{2}(J+iD_{i;j}) S_i^+ S_j^- + h.c. \nn
  &\stackrel{\rm MF}{\longrightarrow}&
  -{1\over 4} (J + iD_{i;j} ) \sum_{\sig}
  \left[ \la \hat{\chi}_{i;j}^{- \sig} \ra \hat{\chi}_{j;i}^{\sig} 
    + \hat{\chi}_{i;j}^{- \sig} \la \hat{\chi}_{j;i}^{\sig} \ra
  \right].
  \nonumber
\end{eqnarray}
Using Eq.\,(5) of main article, we can define the spin current
operator on the kagome lattice, and then obtain
\begin{eqnarray}
  j_{\al}^S = \frac{1}{N_t} \sum_{\triangle_{\v i}} j_{\al}^S (\v i)
  =\sum_{\v k, \sig} \sigma \Psi_{\v k \sig}^{\dag} {\cal S}_{\al \v k \sigma} \Psi_{\v k \sig}, 
  \label{eq:cur_op_spin_H}
\end{eqnarray}
where

\begin{eqnarray}
  {\cal S}_{x \v k \sig} &=& 
  \begin{pmatrix}
    0& 
    \frac{1}{2} \left( J \chi + i \sig D \chi_{-\sig} \right)\sin k_1&
    \frac{1}{2} (J \chi^* - i \sig D \chi_{-\sig}^*) \sin k_3\\
    \frac{1}{2} \left( J \chi^* - i \sig D \chi_{-\sig}^* \right)\sin k_1&
    0&
    (J \chi + i \sig D \chi_{-\sig}) \sin k_2\\
    \frac{1}{2} \left( J \chi + i \sig D \chi_{-\sig} \right)\sin k_3&
    (J \chi^* - i \sig D \chi_{-\sig}^*) \sin k_2&0
  \end{pmatrix},\nn
  {\cal S}_{y \v k \sig} &=& 
  \begin{pmatrix}
    0& 
    \frac{\sqrt{3}}{2} \left( J \chi + i \sig D \chi_{-\sig} \right)\sin k_1&
    -\frac{\sqrt{3}}{2} (J \chi^* - i \sig D \chi_{-\sig}^*) \sin k_3\\
    \frac{\sqrt{3}}{2} \left( J \chi^* - i \sig D \chi_{-\sig}^* \right)\sin k_1&
    0&0\\
    -\frac{\sqrt{3}}{2} \left( J \chi + i \sig D \chi_{-\sig} \right)\sin k_3&
    0&0
\end{pmatrix}.\nonumber
\end{eqnarray}
Again, we find complete equivalence of this to the current operator derived from the mean-field Hamiltonian, 
\begin{eqnarray}
 {\v  j}^S = \sum_{\v k \sigma} \sigma \Psi_{\v k \sig}^{\dag} 
  \frac{\pd H_{\v k \sig}}{\pd \v k} \Psi_{\v k \sig}.
\end{eqnarray}

\end{document}